\documentclass[preprint,showpacs,superscriptaddress,preprintnumbers,amsmath,amssymb,prc]{revtex4-1}

\usepackage{graphicx}
\usepackage{dcolumn}
\usepackage{bm}
\usepackage{ifpdf}
\usepackage{epstopdf}
\usepackage{slashed}
\usepackage{amsfonts}
\usepackage{mathrsfs}
\usepackage{amssymb}
\usepackage{multirow}
\usepackage{CJK}
\usepackage{xcolor}

\begin{document}
\begin{CJK*}{GBK}{song}

\title{Non-relativistic expansion of single-nucleon Dirac equation: Comparison between Foldy-Wouthuysen transformation and similarity renormalization group}

\author{Yixin Guo}
\affiliation{Department of Physics, Graduate School of Science, The University of Tokyo, Tokyo 113-0033, Japan}
\affiliation{RIKEN Nishina Center, Wako 351-0198, Japan}
\affiliation{Department of Modern Physics, University of Science and Technology of China, Hefei 230026, China}

\author{Haozhao Liang}
\email{haozhao.liang@riken.jp}
\affiliation{RIKEN Nishina Center, Wako 351-0198, Japan}
\affiliation{Department of Physics, Graduate School of Science, The University of Tokyo, Tokyo 113-0033, Japan}

\date{\today}

\begin{abstract}
  By following the Foldy-Wouthuysen (FW) transformation of the Dirac equation, we work out the exact analytic expressions up to the $1/M^4$ order for the general cases in the covariant density functional theory.
  These results are further compared with the corresponding ones derived from another novel non-relativistic expansion method, the similarity renormalization group (SRG).
  Based on that, the origin of the difference between the results obtained by the FW transformation and SRG method is explored.
\end{abstract}

\maketitle

\section{Introduction}

The Foldy-Wouthuysen (FW) transformation, also called the Pryce-Tani-Foldy-Wouthuysen transformation \cite{Pryce1948Proc.RoyalSoc.A195.62--81, Foldy1950Phys.Rev.78.29--36, Tani1951Prog.Theor.Phys.6.267--285, Foldy1952Phys.Rev.87.688--693}, was initially formulated around 1950 to describe the non-relativistic limit of the relativistic electrons.
A great advantage of the FW transformation is the simple forms of operators corresponding to classical observables.
This transformation for the spin-$1/2$ particles was soon extended to the cases of the spin-$0$ and spin-$1$ particles \cite{Case1954Phys.Rev.95.1323--1328}, and later generalized to the cases of arbitrary spins \cite{Jayaraman1975J.Phys.AMath.Gen.8.L1--L4}.
Recently, the exact form of the exponential FW operator for an particle with an arbitrary spin was given in Ref.~\cite{Silenko2016Phys.Rev.A94.032104}.

The FW transformation is also an elegant method for the non-relativistic expansion of the Dirac equation \cite{Greiner1990.277--290}.
One milestone of its applications in atomic physics is that, by using the FW transformation in the presence of external fields, the fine structure
\cite{Michelson1887Philos.Mag.andAm.J.Sci.24.463--466, Sommerfeld1940Sci.Nat.28.417--423} describing the splitting of the spectral lines of atoms due to the electron spin and the relativistic corrections to the non-relativistic Schr\"odinger equation was successfully testified \cite{Greiner1990.277--290}.

Due to its elegant and powerful scheme, the FW transformation has been applied in very diverse areas, such as
in electrodynamics \cite{Silenko1995Theor.Math.Phys.105.1224--1230, Silenko2003J.Math.Phys.44.2952--2966},
atomic physics \cite{Hiramoto2001Prog.Theor.Phys.106.1223--1238, Pachucki2005Phys.Rev.A71.012503},
condensed matter physics \cite{Bliokh2005EPL72.7--13, Gosselin2007Eur.Phys.J.B58.137--148, Gosselin2009Eur.Phys.J.C59.883},
optics \cite{Lippert1994EPL27.537--541, Khan2006Optik117.481-488, Hinschberger2012Phys.Lett.A376.813--819},
acoustics \cite{Fishman1992J.Math.Phys.33.1887--1914},
quantum chemistry \cite{Reiher2006Theor.Chem.Acc.116.241--252, Autschbach2007Coord.Chem.Rev.251.1796--1821, Liu2010Mol.Phys.108.1679--1706, Nakajima2012Chem.Rev.112.385--402, Peng2012J.Chem.Phys.136.244108, Peng2012Theor.Chem.Acc.131.1081},
nuclear physics \cite{Arodfmmodecutezlsezi1986Phys.Rev.D33.534--539, Scherer1994Nucl.Phys.A570.686--700},
quantum field theory \cite{Neznamov2006Phys.Part.Nucl.37.86--103},
and gravity \cite{Silenko2005Phys.Rev.D71.064016, Gosselin2011Eur.Phys.J.C71.1739, Obukhov2013Phys.Rev.D88.084014, Obukhov2014Phys.Rev.D90.124068},
in the theory of the weak interaction \cite{Silenko1996Nucl.Instrum.MethodsPhys.Res.Sect.B114.259--262, Silenko1997Theor.Math.Phys.112.922--928},
and even in the scheme of supersymmetric quantum mechanics \cite{Junker2018J.Math.Phys.59.052301}.
In recent years, the FW transformation has also been discussed in a gravitational wave background of the generalized Dirac Hamiltonian in the non-relativistic limit \cite{Goncalves2007Phys.Rev.D75.124023, Quach2015Phys.Rev.D92.084047}.

Nevertheless, in most of the above studies and applications of FW transformation, only the leading-order terms were derived and treated, such as the relativistic correction to the kinetic energy appearing up to the order of $1/M^3$ (with $M$ the bare mass of the particle), and the correction due to the spin-orbit coupling and the Darwin term appearing in the order of $1/M^2$ \cite{Greiner1990.277--290}.
Recently, the FW non-relativistic expansions have been carried out up to the order of $1/M^7$ \cite{Chen2010Phys.Rev.A82.012115} or even $1/M^{14}$ \cite{Chen2014Phys.Rev.A90.012112}.
Nevertheless, such derivations are valid only for a family of potentials with certain properties.

For the systems of atomic nuclei, the covariant density functional theory \cite{Meng2006Prog.Part.Nucl.Phys.57.470--563, Meng2015J.Phys.GNucl.Part.Phys.42.093101, Meng2016.} is one of state-of-the-art methodologies.
During the past decades, it has achieved great success in describing various of nuclear phenomena in both stable and exotic nuclei.
In this theoretical framework, the equation of motion for nucleons is described by the Dirac equation with not only strong vector potential $V(\mathbf{r})$ but also strong scalar potential $S(\mathbf{r})$, whose typical values are $V(\mathbf{r})\sim 300$~MeV and $S(\mathbf{r})\sim -350$~MeV, while the mass of nucleon is $M=939$~MeV.
As a result, the non-relativistic expansion for the single-nucleon Dirac equation is non-trivial, and its high-order terms have to be derived and verified carefully.

In 2012, Guo \cite{Guo2012Phys.Rev.C85.021302} applied the similarity renormalization group (SRG) \cite{Wegner1994Ann.Phys.Berl.506.77--91, Bylev1998Phys.Lett.B428.329--333}, instead of the FW transformation, for the non-relativistic expansion of the single-nucleon Dirac equation up to the ${1}/{M^3}$ order, in order to investigate the nuclear pseudospin symmetry \cite{Liang2015Phys.Rep.570.1--84}.
By using the SRG method, the Dirac Hamiltonian is gradually transformed into a diagonal form with the evolution controlled by the flow equation.
As a result, the eigenequations for the upper and lower components of the Dirac spinors are decoupled at the end of the flow.
Meanwhile, the non-relativistic reduced Hamiltonian thus obtained is Hermitian, and it can also be expanded into the series in terms of ${1}/{M^i}$.
To achieve the accuracy of single-particle energies at the $0.1$~MeV level, very recently, we demonstrated that the  ${1}/{M^4}$-order terms are needed and presented the detailed derivation and results in Ref.~\cite{Guo2019Phys.Rev.C99.054324}.

On the one hand, both the FW transformation and SRG method provide a systematic way to carry out the non-relativistic expansion of Dirac equation up to an arbitrary order.
On the other hand, the FW transformation employs finite steps of unitary transformation, while the SRG method employs infinite infinitesimal unitary transformations along the flow.
Therefore, it is interesting and important to investigate the similarities and differences between these two approaches.

In this paper, we will perform for the first time the exact non-relativistic expansion of the single-nucleon Dirac equation up to the $1/M^4$ order by using the FW transformation.
We will also investigate the difference between these results and those obtained by the SRG method.

This paper is organized as follows. In Section~\ref{sec:SRG}, the SRG method will be first recalled, and the expansion up to the $1/M^4$ order will be shown.
The new expansion up to the same order by using the FW transformation with a general consideration, such as the inclusion of the scalar potential, will be derived in Section~\ref{sec:FW}.
The specific forms of the results under the spherical symmetry will be presented in Section~\ref{IIIA}, and the difference between the SRG and FW results as well as the origin of such a difference will be investigated in Section~\ref{IIIB}.
Finally, a summary and perspectives will be given in Section~\ref{sec:IV}.

\section{Theoretical Framework}\label{sec:II}

\subsection{Non-relativistic expansion by SRG}\label{sec:SRG}

In the theoretical framework of covariant density functional theory \cite{Meng2006Prog.Part.Nucl.Phys.57.470--563, Meng2015J.Phys.GNucl.Part.Phys.42.093101, Meng2016.}, the Dirac Hamiltonian for nucleons reads
\begin{equation}\label{Hamiltonian}
    H = \boldsymbol{\alpha} \cdot \mathbf{p} + \beta (M+S) + V,
\end{equation}
where $\alpha$ and $\beta$ are the Dirac matrices, $M$ is the mass of nucleon, and $S$ and $V$ are the scalar and vector potentials, respectively.
In this section, we recall the technique of SRG \cite{Wegner1994Ann.Phys.Berl.506.77--91, Bylev1998Phys.Lett.B428.329--333} and show the results up to the $1/M^4$ order \cite{Guo2012Phys.Rev.C85.021302, Guo2019Phys.Rev.C99.054324}.

The basic idea of SRG is that the Dirac Hamiltonian in Eq.~(\ref{Hamiltonian}) is transformed by a unitary operator $U(l)$ as
\begin{equation}
H(l)=U(l)HU^\dagger(l),\quad H(0)=H,
\end{equation}
with a flow parameter $l$.
While $U(l)$ is not known explicitly, an anti-Hermitian generator $\eta(l)$ related to $U(l)$ as $\eta(l) = \frac{\textrm{d}U(l)}{\textrm{d}l} U^\dagger(l)$ will be chosen, and the Hamiltonian $H(l)$ evolves with the differential flow equation
\begin{equation}\label{eq:1}
    \frac{\textrm{d}H(l)}{\textrm{d}l}=[\eta(l),H(l)].
\end{equation}
One of convenient and appropriate choices of $\eta(l)$ reads \cite{Bylev1998Phys.Lett.B428.329--333}
\begin{equation}\label{eq:2}
    \eta(l)=[\beta M,H(l)],
\end{equation}
which will eventually transform Dirac Hamiltonian into a diagonal form.

For solving Eq.~(\ref{eq:1}), first of all, the Hamiltonian is divided into two parts:
\begin{equation}\label{eq:3}
    H(l)=\varepsilon(l)+o(l),
\end{equation}
according to the commutation and anti-commutation relations with respect to $\beta$, i.e., $[\varepsilon, \beta] = 0$ and $\{o, \beta\} = 0$.
As a result,
\begin{subequations}
\begin{align}
\frac{\textrm{d}\varepsilon(l)}{\textrm{d}l}&=4M\beta o^2(l),\label{eq:s1} \\
\frac{\textrm{d}o(l)}{\textrm{d}l}&=2M\beta[o(l),\varepsilon(l)],\label{eq:s2}
\end{align}
\end{subequations}
with the initial conditions
\begin{equation}
\varepsilon(0)=\beta(M+S)+V,\qquad o(0)=\boldsymbol{\alpha} \cdot \mathbf{p}.
\end{equation}

Equations~(\ref{eq:s1}) and (\ref{eq:s2}) can then be solved with the expansion into the series of ${1}/{M^i}$ \cite{Bylev1998Phys.Lett.B428.329--333}, which gives
\begin{subequations}
\begin{align}
\frac{\varepsilon(\lambda)}{M}&=\sum^\infty_{i=0}\frac{\varepsilon_i(\lambda)}{M^i},\label{eq:s3}\\
\frac{o(\lambda)}{M}&=\sum^\infty_{j=1}\frac{o_j(\lambda)}{M^j},\label{eq:s4}
\end{align}
\end{subequations}
while the flow parameter is replaced by a dimensionless one $\lambda=lM^2$.
The solutions are obtained as \cite{Guo2012Phys.Rev.C85.021302}
\begin{subequations}
\begin{align}
\varepsilon_n(\lambda) &= \varepsilon_n(0) +4\beta\int_0^\lambda\sum^{n-1}_{k=1}o_k(\lambda')o_{n-k}(\lambda')\,\textrm{d}\lambda',\label{eq:s7}\\
o_n(\lambda) &=o_n(0)e^{-4\lambda} +2\beta e^{-4\lambda}\int_0^\lambda\sum^{n-1}_{k=1}[e^{4\lambda'}o_k(\lambda'),\varepsilon_{n-k}(\lambda')] \,\textrm{d}\lambda',\label{eq:s8}
\end{align}
\end{subequations}
with the initial conditions,
\begin{align}\label{eq:s9}
&\varepsilon_0(0)=\beta,\quad  \varepsilon_1(0)=\beta S+V,\quad  \varepsilon_n(0)=0\quad  \mbox{if}\quad  n\geq2,\nonumber\\
&o_1(0)=\boldsymbol{\alpha} \cdot \mathbf{p},\quad o_n(0)=0\quad \mbox{if}\quad  n\geq2.
\end{align}

Therefore, at the end of the flow $\lambda \rightarrow \infty$, all the off-diagonal parts vanish, i.e., $o_n(\infty) = 0$ for all $n$.
Meanwhile, the diagonalized Dirac operator up to the $1/M^4$ order is not difficult to be obtained as \cite{Guo2019Phys.Rev.C99.054324}
\begin{align}\label{eq:A1}
\mathcal{H}_{\rm SRG}
=&\, \varepsilon(\infty)\nonumber\\
=&\,M\varepsilon_0(\infty)+\varepsilon_1(\infty)+\frac{\varepsilon_2(\infty)}{M}+\frac{\varepsilon_3(\infty)}{M^2}
+\frac{\varepsilon_4(\infty)}{M^3}+\frac{\varepsilon_5(\infty)}{M^4}+\cdots\nonumber\\
=&\,M\varepsilon_0(0)+\varepsilon_1(0)+\frac{1}{2M}\beta o_1^2(0)
+\frac{1}{8M^2}[[o_1(0),\varepsilon_1(0)],o_1(0)]\nonumber\\
&+\frac{1}{32M^3}\beta\bigg(-4o_1^4(0)+[[o_1(0),\varepsilon_1(0)],\varepsilon_1(0)]o_1(0)\nonumber\\
&\qquad+o_1(0)[[o_1(0),\varepsilon_1(0)],\varepsilon_1(0)] -2[o_1(0),\varepsilon_1(0)][o_1(0),\varepsilon_1(0)]\bigg)\nonumber\\
&+\frac{1}{128M^4}\bigg(-9[[o_1(0),\varepsilon_1(0)],o_1^3(0)] +3[o_1(0),\varepsilon_1(0)]^2\varepsilon_1(0)\nonumber\\ &\qquad+3\varepsilon_1(0)[o_1(0),\varepsilon_1(0)]^2
-6[o_1(0),\varepsilon_1(0)]\varepsilon_1(0)[o_1(0),\varepsilon_1(0)]\nonumber\\
&\qquad+3[o_1(0)[o_1(0),\varepsilon_1(0)]o_1(0),o_1(0)
+[[[[o_1(0),\varepsilon_1(0)],\varepsilon_1(0)],\varepsilon_1(0)],o_1(0)]\bigg)\nonumber\\
&+\cdots
\end{align}

\subsection{Non-relativistic expansion by FW transformation}\label{sec:FW}

In this section, we perform for the first time the exact non-relativistic expansion of the single-nucleon Dirac equation up to the $1/M^4$ order by using the FW transformation for the general cases, in particular when the scalar potential is considered.

According to the FW transformation in the presence of external fields \cite{Foldy1950Phys.Rev.78.29--36, Foldy1952Phys.Rev.87.688--693,Greiner1990.277--290}, the corresponding operators are defined as
\begin{align}
          O=\boldsymbol{\alpha} \cdot \mathbf{p},\qquad
\varepsilon=\beta S+V,\qquad
    \Lambda=-\frac{i}{2M}\beta O,
\end{align}
where the operators $O$ and $\varepsilon$ satisfy $O \beta=-\beta O$ and $\varepsilon \beta = \beta \varepsilon$, respectively.
The Hamiltonian~\eqref{Hamiltonian} reads
\begin{align}
H = \beta M+O+\varepsilon.
\end{align}

Based on the above definitions, the Dirac Hamiltonian is transformed by a unitary operation into \cite{Greiner1990.277--290}
\begin{align}
H'=\,&e^{i\Lambda}He^{-i\Lambda}\nonumber\\
           =\,&H+i[\Lambda,H] +\frac{i^2}{2!}[\Lambda,[\Lambda,H]] +\cdots 
           +\frac{i^n}{n!}[\underbrace{\Lambda,[\Lambda,\cdots,[\Lambda}_n,H]\cdots]] +\cdots
\end{align}
where
\begin{align}
\frac{i^n}{n!}[\underbrace{\Lambda,[\Lambda,\cdots,[\Lambda}_n,H]\cdots]]
             =(-1)^{\frac{n(n-1)}{2}}\frac{\beta^n}{n!M^n}(O^n\beta M+O^{n+1}
             +\frac{1}{2^n}[\underbrace{O,[O,\cdots,[O}_n,\varepsilon]\cdots]]).
\end{align}
Keeping all the terms up to the $1/M^n$ order, the unitary transformed Hamiltonian reads
\begin{align}
H'_{1/M^n}
=\,&\beta M+\varepsilon+\sum_{k=1}^{n+1}(-1)^{\frac{(k-1)(k-2)}{2}}\frac{k-1}{k!M^{k-1}}\beta^{k-1}O^{k}\nonumber\\
&+\sum_{k=0}^n(-1)^{\frac{k(k-1)}{2}}\frac{\beta^k}{2^kk!M^k}[\underbrace{O,[O,\cdots,[O}_k,\varepsilon]\cdots]].
\end{align}
For example, the corresponding result up to the $1/M^4$ order is
\begin{align}\label{firstH}
H'_{1/M^4}
=\,&\beta M +\varepsilon +\frac{1}{2M}\beta O^2 -\frac{1}{3M^2}O^3 -\frac{1}{8M^3}\beta O^4 +\frac{1}{30M^4}O^5\nonumber\\
   &+\frac{1}{2M}\beta[O,\varepsilon] -\frac{1}{8M^2}[O,[O,\varepsilon]]
   -\frac{1}{48M^3}\beta[O,[O,[O,\varepsilon]]]
   +\frac{1}{384M^4}[O,[O,[O,[O,\varepsilon]]]].
\end{align}

In this unitary transformed Hamiltonian, e.g., Eq.~(\ref{firstH}), the off-diagonal parts are not zero but raised by one order from $O$ to $\frac{1}{2M}\beta[O,\varepsilon]$ (plus higher-order terms).
In order to make the off-diagonal parts vanish, more precisely speaking, to make them higher than a given order, one should repeat the FW transformation until the accuracy is achieved \cite{Greiner1990.277--290}.
For that, the operators $O$ and $\varepsilon$ can be redefined according to Eq.~(\ref{firstH}), and one has
\begin{subequations}
\begin{align}
\varepsilon' = \,&\varepsilon +\frac{1}{2M}\beta O^2 -\frac{1}{8M^2}[O,[O,\varepsilon]] -\frac{1}{8M^3}\beta O^4 
            +\frac{1}{384M^4}[O,[O,[O,[O,\varepsilon]]]],\\
          O' = \,&\frac{1}{2M}\beta[O,\varepsilon] -\frac{1}{3M^2}O^3 -\frac{1}{48M^3}\beta[O,[O,[O,\varepsilon]]]
          +\frac{1}{30M^4}O^5,\\
    \Lambda' = \,&-\frac{i}{2M}\beta O'.
\end{align}
\end{subequations}
The corresponding FW transformation reads
\begin{align}
H''= e^{i\Lambda'} H' e^{-i\Lambda'}.
\end{align}
It can be seen that in $H''$ the off-diagonal parts are of the order of $1/M^2$.
Therefore, one should repeat this procedure to $H'''''$ to make the off-diagonal parts of the order of $1/M^5$.
Of course, to get the resultant diagonal parts in $H'''''$ up to the $1/M^4$ order, a recursion technique makes the calculation less complicated than it looks.

As a result, the non-relativistic expansion by the FW transformation up to the $1/M^4$ order reads
\begin{align}\label{FWH}
\mathcal{H}_{\rm FW}
=\,&\beta M +\varepsilon +\frac{1}{2M}\beta O^2 -\frac{1}{8M^2}[O,[O,\varepsilon]]-\frac{1}{8M^3}\beta O^4 -\frac{1}{8M^3}\beta[O,\varepsilon][O,\varepsilon] \nonumber\\
& 
+\frac{1}{384M^4}[O,[O,[O,[O,\varepsilon]]]] +\frac{1}{12M^4}[O^3,[O,\varepsilon]]  \nonumber\\
&
+\frac{1}{32M^4}[O,\varepsilon][O,\varepsilon]\varepsilon
-\frac{1}{16M^4}[O,\varepsilon]\varepsilon[O,\varepsilon]
+\frac{1}{32M^4}\varepsilon[O,\varepsilon][O,\varepsilon].
\end{align}

\section{Results and Discussion}\label{sec:III}

\subsection{Results with spherical symmetry}\label{IIIA}

For the systems with spherical symmetry, i.e., for spherical nuclei, the corresponding radial single-nucleon Dirac equation reads \cite{Meng2006Prog.Part.Nucl.Phys.57.470--563, Meng2015J.Phys.GNucl.Part.Phys.42.093101, Meng2016.}
\begin{equation}\label{eq:Dirac1}
\left(
\begin{array}{cc}
\Sigma(r)+M & -\frac{\textrm{d}}{\textrm{d}r}+\frac{\kappa}{r}\\
\frac{\textrm{d}}{\textrm{d}r}+\frac{\kappa}{r} & \Delta(r)-M
\end{array}
\right )
\left(
\begin{array}{c}
G(r) \\ F(r)
\end{array}
\right )
=E
\left(
\begin{array}{c}
G(r) \\ F(r)
\end{array}
\right ),
\end{equation}
where $\kappa$ is a good quantum number defined as $\kappa=\mp~(j+{1}/{2})$ for $j=l\pm{1}/{2}$, and $\Sigma(r) = V(r) + S(r)$ and $\Delta(r) = V(r) - S(r)$ are the sum of and the difference between the vector and scalar potentials, respectively.
The single-particle energies $E = \varepsilon +M $ include the mass of nucleon.
The operators $\varepsilon$ and $O$ read
\begin{equation}
\varepsilon={\left(
\begin{array}{cc}
\Sigma(r) & 0\\
0 & \Delta(r)
\end{array}
\right )},\quad
O={\left(
\begin{array}{cc}
0 &-\frac{\textrm{d}}{\textrm{d}r}+\frac{\kappa}{r}\\
\frac{\textrm{d}}{\textrm{d}r}+\frac{\kappa}{r}& 0
\end{array}
\right )}.
\end{equation}

According to Eq.~(\ref{FWH}), the Dirac Hamiltonian is transformed by the FW transformation as
\begin{equation}\label{eq:HFW}
{\left( \begin{array}{cc}
\mathcal{H}^{\rm (F)}_{\rm FW} + M & O(\frac{1}{M^5}) \\
O(\frac{1}{M^5}) & \mathcal{H}^{\rm (D)}_{\rm FW} - M
\end{array}
\right)}.
\end{equation}
It is seen that the off-diagonal parts are not strictly zero, which is different from the results of SRG, but they are of higher order than the required one.
Focusing on the single-particle states in the Fermi sea, which correspond to their counterparts in the non-relativistic framework, the explicit expansions of $\mathcal{H}^{\rm (F)}_{\rm FW}$ up to the $1/M^4$ order is carefully worked out as
\begin{subequations}
\begin{align}
\mathcal{H}^{\rm (F)}_{0,\textrm{FW}}=\,&\Sigma(r),\label{eq1}\\
\mathcal{H}^{\rm (F)}_{1,\textrm{FW}}=\,&\frac{1}{2M}p^2,\label{eq2}\\
\mathcal{H}^{\rm (F)}_{2,\textrm{FW}}=\,&\frac{1}{8M^2}\bigg(-4Sp^2 +4S'\frac{\textrm{d}}{\textrm{d}r} -2\frac{\kappa}{r}\Delta' +\Sigma''\bigg),\label{eq3}\\
\mathcal{H}^{\rm (F)}_{3,\textrm{FW}}=\,&\frac{1}{8M^3}\bigg(-p^4 +4S^2p^2 -8SS'\frac{\textrm{d}}{\textrm{d}r}-2S\Sigma''
+4S\Delta'\frac{\kappa}{r} +\Sigma'\Delta'\bigg),\label{eq4}\\
\mathcal{H}^{\rm (F)}_{4,\textrm{FW}}=\,&\frac{1}{384M^4}\bigg\{144Sp^4 -288S'p^2\frac{\textrm{d}}{\textrm{d}r}
                +\bigg[72\Delta'\frac{\kappa}{r}-24(4\Sigma''+9S'')-192S^3\bigg]p^2\nonumber\\
               &+\bigg[72\Delta'\frac{\kappa}{r^2} -72\Delta''\frac{\kappa}{r} +24(3S'''+4\Sigma''') +576S^2S'\bigg]\frac{\textrm{d}}{\textrm{d}r}\nonumber\\
               & +\bigg[-12(5\Sigma'-36S')\frac{\kappa(\kappa+1)}{r^3}
               +12(5\Sigma''+12S'')\frac{\kappa(\kappa+1)}{r^2} -72\Delta'\frac{\kappa}{r^3} +72\Delta''\frac{\kappa}{r^2} \nonumber\\
               &-36\Delta'''\frac{\kappa}{r} -288S^2\Delta'\frac{\kappa}{r} +33\Sigma'''' +144S^2\Sigma''
               +24S\Sigma'(2\Sigma'-6\Delta')\bigg]\bigg\},\label{eq5}
\end{align}
\end{subequations}
where
\begin{equation}
p^2=-\frac{\textrm{d}^2}{\textrm{d}r^2}+\frac{\kappa(\kappa+1)}{r^2},
\end{equation}
and
\begin{align}
p^4= \frac{\textrm{d}^4}{\textrm{d}r^4} -2\frac{\kappa(\kappa+1)}{r^2}\frac{\textrm{d}^2}{\textrm{d}r^2} +4\frac{\kappa(\kappa+1)}{r^3}\frac{\textrm{d}}{\textrm{d}r} +\frac{\kappa(\kappa+1)(\kappa+3)(\kappa-2)}{r^4}.
\end{align}

In contrast, the Dirac Hamiltonian transformed by the SRG method reads
\begin{equation}\label{eq:HSRG}
{\left( \begin{array}{cc}
\mathcal{H}^{\rm (F)}_{\rm SRG} + M & 0 \\
0 & \mathcal{H}^{\rm (D)}_{\rm SRG} - M
\end{array}
\right)}.
\end{equation}
Its off-diagonal parts strictly vanish up to infinite order.
According to Eqs.~\eqref{eq:A1}, the explicit expansions of $\mathcal{H}^{\rm (F)}_{\rm SRG}$ up to the $1/M^4$ order reads \cite{Guo2019Phys.Rev.C99.054324}
\begin{subequations}
\begin{align}
\mathcal{H}^{\rm (F)}_{0,\textrm{SRG}}=&\Sigma(r),\label{eq:p0}\\
\mathcal{H}^{\rm (F)}_{1,\textrm{SRG}}=&\frac{1}{2M}p^2,\label{eq:p1}\\
\mathcal{H}^{\rm (F)}_{2,\textrm{SRG}}=&\frac{1}{8M^2}\bigg(-4Sp^2 +4S'\frac{\textrm{d}}{\textrm{d}r}
            -2\frac{\kappa}{r}\Delta' +\Sigma''\bigg),\label{eq:p2}\\
\mathcal{H}^{\rm (F)}_{3,\textrm{SRG}}=&\frac{1}{32M^3}\bigg(-4p^4 +16S^2p^2-32SS'\frac{\textrm{d}}{\textrm{d}r}
            -8S\Sigma'' +16S\Delta'\frac{\kappa}{r} -2\Sigma'^2 +4\Sigma'\Delta'\bigg),\label{eq:p3}\\
\mathcal{H}^{\rm (F)}_{4,\textrm{SRG}}=&\frac{1}{128M^4}
            \bigg\{48Sp^4 -96S'p^2\frac{\textrm{d}}{\textrm{d}r}
            +\bigg[24\Delta'\frac{\kappa}{r} -24(\Sigma''+3S'')-64S^3\bigg]p^2\nonumber\\
            &+\bigg[24\Delta'\frac{\kappa}{r^2} -24\Delta''\frac{\kappa}{r} +24(S'''+\Sigma''') +192S^2S'\bigg]\frac{\textrm{d}}{\textrm{d}r}\nonumber\\
            & +\bigg[-12(\Sigma'-12S')\frac{\kappa(\kappa+1)}{r^3} +12(\Sigma''+4S'')\frac{\kappa(\kappa+1)}{r^2} -24\Delta'\frac{\kappa}{r^3} +24\Delta''\frac{\kappa}{r^2} \nonumber\\
            & -12\Delta'''\frac{\kappa}{r} -96S^2\Delta'\frac{\kappa}{r} +9\Sigma''''
            +48S^2\Sigma''+24S\Sigma'(\Sigma'-2\Delta')\bigg]\bigg\}.\label{eq:p4}
\end{align}
\end{subequations}

\subsection{Comparison between FW and SRG approaches}\label{IIIB}

By comparing Eqs.~\eqref{eq:A1} and \eqref{FWH}, it is found that
\begin{subequations}\label{eq:diff}
\begin{align}
    \mathcal{H}_{0,{\rm SRG}} - \mathcal{H}_{0,{\rm FW}} = \,&0,\\
    \mathcal{H}_{1,{\rm SRG}} - \mathcal{H}_{1,{\rm FW}} = \,&0,\\
    \mathcal{H}_{2,{\rm SRG}} - \mathcal{H}_{2,{\rm FW}} =\,& 0,\\
    \mathcal{H}_{3,{\rm SRG}} - \mathcal{H}_{3,{\rm FW}} = \,&\frac{\beta}{32M^3}[[O^2,\varepsilon], \varepsilon],\\
    \mathcal{H}_{4,{\rm SRG}} - \mathcal{H}_{4,{\rm FW}} =\,& \frac{1}{64M^4}[[O^2,\varepsilon], O^2]
        +\frac{1}{128M^4}\left[\left([\varepsilon^2,O^2]-2[\varepsilon, O\varepsilon O]\right),\varepsilon\right].
\end{align}
\end{subequations}
Such results seem to indicate that the FW transformation and SRG method agree with each other up to the $1/M^2$ order, but they lead to different results starting from the $1/M^3$ order.
However, considering the infinite mass limit $M \rightarrow \infty$, all the expressions should be strictly organized order by order in this limit, and thus the difference between Eqs.~\eqref{eq:HSRG} and \eqref{eq:HFW} in the $1/M^5$ order cannot lead to the differences of the results in the $1/M^3$ and $1/M^4$ orders.
This puzzle is eager to be investigated.

It turns out that the differences shown in Eq.~\eqref{eq:diff} comes from an additional unitary transformation, after the Dirac Hamiltonian is decoupled into the upper and lower parts.
Let
\begin{equation}
  \Xi = -\frac{i\beta}{32M^3}[O^2, \varepsilon] - \frac{i}{128M^4}\left([\varepsilon^2,O^2]-2[\varepsilon, O\varepsilon O]\right).
\end{equation}
It is a Hermitian and diagonal operator, i.e., $\Xi^\dag = \Xi$ and $\beta\Xi = \Xi\beta$.
Acting an additional unitary transformation on $\mathcal{H}_{\rm FW}$, it reads
\begin{equation}
  e^{i\Xi}\, \mathcal{H}_{\rm FW}\, e^{-i\Xi} = \mathcal{H}_{\rm FW} + i [\Xi, \mathcal{H}_{\rm FW}] + \cdots
\end{equation}
Keeping all the terms up to the $1/M^4$ order, the results are
\begin{align}
  & e^{i\Xi}\, \mathcal{H}_{\rm FW}\, e^{-i\Xi}\nonumber\\
   =&\, \mathcal{H}_{\rm FW} + \frac{\beta}{32M^3}[[O^2,\varepsilon], \varepsilon]
    + \frac{1}{64M^4}[[O^2,\varepsilon], O^2]
    +\frac{1}{128M^4}\left[\left([\varepsilon^2,O^2]-2[\varepsilon, O\varepsilon O]\right),\varepsilon\right],
\end{align}
which is nothing but $\mathcal{H}_{\rm SRG}$.

In the spherical case, the explicit expressions of operators $\frac{\beta}{32M^3}[[O^2,\varepsilon], \varepsilon]$, $\frac{1}{64M^4}[[O^2,\varepsilon], O^2]$, and $\frac{1}{128M^4}\left[\left([\varepsilon^2,O^2]-2[\varepsilon, O\varepsilon O]\right),\varepsilon\right]$ acting on the single-particle states in the Fermi sea read
\begin{subequations}
\begin{align}
    -\frac{1}{16M^3} &{\Sigma'}^2,\\
    \frac{1}{64M^4}&\bigg[4\Sigma'' p^2 - 4\Sigma'''\frac{d}{dr} + 4\Sigma'\frac{\kappa(\kappa+1)}{r^3}
        - 4\Sigma''\frac{\kappa(\kappa+1)}{r^2} - \Sigma'''' \bigg],\\
    \frac{1}{16M^4}&S {\Sigma'}^2,
\end{align}
\end{subequations}
respectively.
This answers the differences shown between Eqs.~(\ref{eq4}) and (\ref{eq:p3}) as well as between Eqs.~(\ref{eq5}) and (\ref{eq:p4}).

Since $e^{-i\Xi}$ is an additional unitary transformation acting on the already decoupled Hamiltonian, it does not affect the non-relativistic expansion of the Dirac equation, and the single-particle spectra obtained by the FW transformation and the SRG method will be the same to each other.

\section{Summary and Perspectives}\label{sec:IV}

In this work, we investigate the non-relativistic expansion of the single-nucleon Dirac equation for general cases in the theoretical framework of covariant density functional theory, where the scalar potential is included.
We work out the exact analytical expansions up to the $1/M^4$ order for the first time by following the FW transformation.
With a further investigation of the difference between Eqs.~\eqref{eq:HSRG} and \eqref{eq:HFW}, i.e., the puzzle that the disagreement seems to appear between the results obtained by the FW transformation and the SRG method, has been well justified.
In other words, the non-relativistic expansion of the Dirac equation is affected, but the single-particle spectra obtained by the FW transformation and the SRG method are the same to each other.

Since the non-relativistic expansion of the Dirac equation by the FW transformation has been extended to needed high orders for nuclear systems.
Similarly to the SRG method, we anticipate that this novel non-relativistic expansion method will establish a potential bridge between the relativistic and non-relativistic density functional theories for the future studies.
In particular, with the present FW transformation, all the unitary transformations involved hold explicit forms.
This leads straightforwardly to the corresponding transformation on other relevant operators in the framework, such as the one-body density operators, and so on.

\begin{acknowledgments}

This work was partially supported by the JSPS Grant-in-Aid for Early-Career Scientists under Grant No.~18K13549 and the JSPS-NSFC Bilateral Program for Joint Research Project on Nuclear mass and life for unravelling mysteries of the $r$-process.

\end{acknowledgments}

\end{CJK*}
\end{document}